\documentclass[review]{elsarticle}

\usepackage{amssymb}
\usepackage{graphics}
\usepackage{graphicx}
\usepackage{epstopdf}
\usepackage{amsmath}
\usepackage{multirow}
\usepackage{bm}
\usepackage{enumerate}
\usepackage{caption, subcaption}
\usepackage{verbatim}
\usepackage{algorithmicx}
\usepackage{algpseudocode}
\usepackage{algorithm}
\usepackage{verbatim}
\usepackage{dsfont}
\usepackage{pbox}

\usepackage[english]{babel} 

\begin{document}
\title{Algorithms for the Construction of Incoherent Frames Under Various Design Constraints}

\author[cr]{Cristian Rusu}
\ead{cristian.rusu@ncirl.ie}
\author[rvt]{Nuria Gonz\'{a}lez-Prelcic}
\ead{nuria@gts.uvigo.es}
\author[rwh]{Robert W. Heath Jr.}
\ead{rheath@utexas.edu}

\address[cr]{The National College of Ireland, Ireland}
\address[rvt]{The University of Vigo, Spain}
\address[rwh]{The University of Texas at Austin, Austin, TX 78712, USA}

\cortext[cor1]{This work was partially funded by the Agencia Estatal de Investigaci\'{o}n (Spain) and the European Regional Development Fund (ERDF) under project MYRADA (TEC2016-75103-C2-2-R), by the Xunta de Galicia (Agrupaci\'{o}n Estrat\'{e}xica Consolidada de Galicia accreditation 2016-2019; Red Temática RedTEIC 2017-2018), and  by the National Science Foundation under Grant No. 1711702.
\newline	
Matlab$^\text{\textcopyright}$ source code is available online https://github.com/cristian-rusu-research/SIDCO.}

\begin{abstract}
Unit norm finite frames are generalizations of orthonormal bases with many applications in signal processing. An important property of a frame is its coherence, a measure of how close any two vectors of the frame are to each other. Low coherence frames are useful in compressed sensing applications. When used as measurement matrices, they successfully recover highly sparse solutions to linear inverse problems. This paper describes algorithms for the design of various low coherence frame types: real, complex, unital (constant magnitude) complex, sparse real and complex, nonnegative real and complex, and harmonic (selection of rows from Fourier matrices). The proposed methods are based on solving a sequence of convex optimization problems that update each vector of the frame. This update reduces the coherence with the other frame vectors, while other constraints on its entries are also imposed. Numerical experiments show the effectiveness of the methods compared to the Welch bound, as well as other competing algorithms, in compressed sensing applications.
\end{abstract}


\maketitle

\newtheorem{remark}{Remark}[section]
\newtheorem{definition}{Definition}[section]
\newtheorem{result}{Result}[section]
\newtheorem{example}{Example}[section]

\section{Introduction}

A frame in finite dimensions \cite{FirstFrames1952, FiniteFramesBook2013} is just a collection of vectors that span the whole space (usually redundant or overcomplete -- meaning more vectors than the dimension of the working space). For a frame of $N$ vectors in $\mathbb{C}^m$, the redundancy is defined as the ratio $\rho = N/m$. The redundancy of a frame defines one of its key properties, as it directly affects several design objectives such as the resilience to erasures, additive noise and quantization \cite{GrassmannianApplication1}. Overcomplete frames have been used with great success in the time-frequency analysis/synthesis \cite{TFAnalysis} of natural signals (Gabor \cite{Gabor},\cite{RecentGabor} and wavelet \cite{Wavelet} transforms, including adaptive representations \cite{AA}) while publicly available toolboxes that efficiently implement these techniques \cite{AnotherToolbox},\cite{ltfatnote015} have eased their use in signal processing applications.

Incoherent frames, frames whose vectors have large distances (angles) between them, have applications in coding and communications \cite{GrassmannianApplication1}, as well as in sparse representations/compressive sensing \cite[Chapter 5]{SparseRepCS}. In compressed sensing, the success of the sparse approximation algorithms based on greedy methods \cite{GreedIsGood} and $\ell_1$ relaxation \cite{JustRelax}, depends upon the incoherence of the measurement dictionaries (frames) used. These theoretical developments show that, in the deterministic case, the success of recovery algorithms to return the true sparse solution depends on the incoherence of the measurement frames: the lower the coherence the higher the success rate of the recovery methods as a function of sparsity. The mutual coherence is also connected to the restricted isometry property (RIP) concept \cite[Chapter 6]{SparseRepCS}, that provides guarantees for sparse recovery in the case where random measurements are performed.

Due to their potential applications, it is valuable to design highly incoherent frames in any $m$-dimensional vector space and with any number of unit norm vectors $N>m$. In many applications, the unit norm constraint is not the only constraint that needs to be imposed on the frame. For example, unital incoherent frames (whose entries have the same magnitude), or incoherent frames constructed by selecting rows from Fourier matrices (which are also unital), are important in communication applications \cite{GrassmannianApplication1}\cite[Chapter~7]{FiniteFramesBook2013}. In this case, the unital constraints reflect practical limitations of the communications hardware. Also, sparse incoherent frames reduce in general the computational complexity of using the frames as analysis/synthesis operators \cite{SparseFrames2011, SparseFrames2014}, while nonnegative incoherent frames are useful in nonnegative matrix factorizations \cite{NNMF2001}. Therefore, although in many signal processing applications \cite{FiniteFramesBook2013} incoherence is a desirable property, there are hard, real-world system constraints that are more important than the incoherence requirement.

For certain dimensions, it is possible to construct tight optimally incoherent \cite{WB} real, complex or unital/harmonic \cite{ETF2, DifferenceSets} frames called equiangular tight frames (ETFs). Their construction, which is primarily based on tools from number and graph theory, is elegant, but unfortunately only possible for few values of $m$, $N$. Thus, it is convenient to have a numerical procedure that can design highly incoherent frames (without the guarantee that these are in fact the most incoherent) for any $m$ and $N$.

There is extensive prior work on the creation of incoherent frames, either real or complex valued, not necessarily equiangular, by iterative optimization \cite{AlterningProjection} \cite{Dhilon}. Most of these numerical algorithms typically deploy some alternating minimization, for a fixed number of iterations, that lowers the coherence. This minimization process is not in general, however, monotonically convergent to a local minimum. Some methods focus on reducing the magnitudes of the off-diagonal entries of the Gram matrix of these frames, followed by a factorization step based on the singular value decomposition \cite{AlterningProjection, AlterningProjectionCode} to recover the frame vectors. Because these algorithms do not deal directly with the frame vectors, but with the Gram matrix of the frame, they are faced with an important limitation: it is very difficult, or computationally expensive, to impose additional constraints to the entries of the vectors. The main reason behind this limitation is that the singular value decomposition (used to factor out the frame synthesis matrix from the Gram matrix) does not easily accommodate for additional constrains in the factorization. A recent new development based on convex optimization, deals with the frame entries directly, and has proven very efficient for the design of incoherent real valued frames \cite{IDCO, OurPaper,Sadeghi2017,BDIDCO}. An advantage of this method is that the addition of new constraints that are convex, or that have an efficient convex relaxation, is relatively simple.

In this paper, we provide designs for both real and complex frames, and consider various specific constraints: non-negativity, sparsity and unit magnitude. We describe the design problems as general optimization problems, and then provide tractable solutions via convex relaxation. These formulations allow us to reduce coherence by establishing a reference frame and a trust region around this frame, where to search for a new, more incoherent, frame. The proposed work extends and improves upon previous methods that are not able to impose further constraints to the frame \cite{AlterningProjection, Dhilon, OurPaper}.

In the special case of designing incoherent harmonic frames (selections of rows from Fourier matrices), due to the distinct nature of the problem, we introduce a different type of convex optimization problem. Our approach is based on iterative reweighted $\ell_\infty / \ell_1$ norm minimization, which is a relaxation of a binary optimization problem that we formulate. This method naturally extends also to the design of incoherent Hadamard frames, selections of rows from Hadamard matrices, and in general to selections from any unital matrix.

We use our proposed algorithms to compute incoherent frames of various dimensions. We compare the impact of adding the different constraints on the coherence, and we show that the complex valued incoherent frames approach the lower bound of performance in many cases. In the harmonic and Hadamard cases, we show that the proposed method matches the best results in all the cases where an exhaustive search was computationally possible (low $m$ and $N$). We also compare with previously proposed methods, and show that our algorithms match or outperform them. Finally, we present a compressive sensing application where the incoherent frames designed by the proposed methods improve the sparse approximation performance of the OMP algorithm \cite{GreedIsGood}. Since incoherent frame design is mostly performed off-line in general, because it depends only on sizes $(m, N)$, we do not focus on the computational complexity analysis of the algorithms. Still, the convex optimization \cite{CO} problems we propose are solved in polynomial time in $m$ and $N$ by standard solvers \cite{CVX}.

\section{Frames background}

A finite unit-norm frame over the field $\mathbb{C}^m$ is a sequence of $N$ vectors $\mathbf{f}_i \in \mathbb{C}^m, 1 \leq i \leq N,$ that satisfies $\| \mathbf{f}_i \|_2 = 1$ and there exists constants $0 < \alpha \leq \beta < \infty$, called the lower and upper bounds of the frame, such that
$	\alpha \|\mathbf{v}\|_2^{2} \leq \sum_{i=1}^{N} |\mathbf{f}_i^H \mathbf{v}|^2 \leq \beta \|\mathbf{v}\|_2^{2},\ \forall \ \mathbf{v} \in \mathbb{C}^m.$
We introduce the frame synthesis matrix $\mathbf{F} \in \mathbb{C}^{m \times N}$ consisting of the concatenated frame vectors $\mathbf{F} = \begin{bmatrix} \mathbf{f}_1 & \mathbf{f}_2 & \dots & \mathbf{f}_N \end{bmatrix}$, i.e., $\mathbf{Fc} = \sum_{i=1}^N c_i \mathbf{f}_i$ \cite[Chapter~1]{FiniteFramesBook2013}. If $\alpha = \beta$ then the frame is called tight and it enjoys excellent properties in terms of reconstruction accuracy \cite{ControlledFrames}.

The mutual coherence \cite[Chapters~6.7,~6.8~and~9.1]{FiniteFramesBook2013} of $\mathbf{F}$ with and without normalization is given by
\begin{equation}
	\mu_0 (\mathbf{F}) = \underset{1\leq i < j \leq N}{\max}{ \frac{ |\mathbf{f}_i^H \mathbf{f}_j| }  {\| \mathbf{f}_i \|_2 \| \mathbf{f}_j \|_2}}, \ 	\mu (\mathbf{F}) = \underset{1\leq i < j \leq N}{\max}{ |\mathbf{f}_i^H \mathbf{f}_j| }.
	\label{eq:MutualCoherence1}
\end{equation}
If the vectors of the frames are normalized, i.e., $\| \mathbf{f}_i \|_2 = 1$ for all $i$, then $\mu_0(\mathbf{F})$ and $\mu(\mathbf{F})$ are identical. If the vectors of the frame have the same $\ell_2$ norm, i.e., $\| \mathbf{f}_i \|_2 = \gamma$ for all $i$, then $\mu_0 (\mathbf{F}) = \gamma^{-2} \mu(\mathbf{F})$; notice that the maximum occurs in those cases for the same indices $(i,j)$ for both definitions in \eqref{eq:MutualCoherence1}. Consider for example $N \times N$ complex unital matrices, i.e., matrices with unit (or constant) magnitude entries. If we select any $m$ rows from such a matrix, then each column has the same energy $\left( \gamma =  \sqrt{m} \right)$, and thus we need not worry about the normalization step of the computation of the mutual coherence if the goal is to minimize it. As shown in the next sections, this observation allows for the introduction of an algorithm for the selection of a given number of rows from a fixed unital matrix, such that the mutual coherence of the extracted $m \times N$ frame is greatly reduced. If we select any $m$ rows from a matrix which is not unital, then the normalization operations in \eqref{eq:MutualCoherence1} become essential for the correct calculation of the coherence, and thus can no longer be avoided.

The Gram matrix \cite[Chapters 1.4.3 and 6.7]{FiniteFramesBook2013} of the normalized full rank frame $\mathbf{F}$ is $\mathbf{G} = \mathbf{F}^H \mathbf{F} \in \mathbb{C}^{N \times N}$. This matrix is Hermitian and positive semidefinite with rank $m$ and it has a unit diagonal (because of the normalized frame vectors). Therefore, only $m$ of its $N$ eigenvalues $\lambda_i$ are nonzero and they obey $\sum_{i=1}^m \lambda_i = \text{tr}(\mathbf{G}) = N$ and $\sum_{i=1}^m \lambda_i^2 = \text{tr}(\mathbf{G}^H \mathbf{G}) \geq m^{-1}N^2$ (in fact, if $\mathbf{F}$ is a tight frame then $\lambda_i = m^{-1}N$ for $1 \leq i \leq m$) \cite[Section~2]{ETF2}. Also, the off-diagonal entries are equal to the inner products between any two distinct frame vectors. The mutual coherence of the frame $\mathbf{F}$ is
\begin{equation}
	\mu (\mathbf{F}) = \underset{1\leq i < j \leq N}{\max}{|g_{ij}|},
\end{equation}
i.e., the maximum absolute value of the off-diagonal entries of the Gram matrix $\mathbf{G}$. The lowest value of the mutual coherence, called the Welch bound (WB) \cite{WB}, when $N \leq m^2$ for complex valued frames is
\begin{equation}
	\mu = \sqrt{\frac{N - m}{m(N-1)}}.
\end{equation}
This bound, which is discussed in detail in \cite[Chapter 5.2]{SparseRepCS}, is achieved by equiangular tight frames (in this case all off-diagonal entries of the Gram matrix are equal in absolute value to the Welch bound) \cite[Section~2]{ETF2}. Therefore, when an equiangular tight frame exists, it is known to be the most incoherent. Sadly, these frames exist only for few pairs $(m,N)$. For any $(m,N)$, the frame, not necessarily tight, that achieves the minimum mutual coherence is called Grassmannian \cite{GrassmannianApplication1}. Denote the squared Frobenius norm of $\mathbf{F} \in \mathbb{C}^{m \times N}$ as $\| \mathbf{F} \|_F^2 = \text{tr}(\mathbf{F}^H \mathbf{F}) = \sum_{i=1}^m \sum_{j=1}^N |f_{ij}|^2$. Then, an important and well known result from the frame literature, that we often use in this manuscript, is the following:

\textbf{Theorem 2 of \cite{AlterningProjection}.} Let $\mathbf{F}$ be a $m \times N$ full rank matrix with singular value decomposition $\mathbf{U\Sigma V}^H$. With respect to the Frobenius norm, the closest $\alpha$-tight frame to $\mathbf{F}$ is given by $\alpha\mathbf{UV}^H$.\hfill$\blacksquare$

\section{Real valued incoherent frame design}

Recently, a method for the design of real valued incoherent frames $(m, N)$ was introduced \cite{OurPaper}. The method is based on convex optimization used in an iterative fashion: each frame vector is updated such that its coherence with the other frame vectors, which are fixed, is reduced. Herein, the previously introduced SIDCO method for real valued frames from \cite{IDCO, OurPaper} and shown in Algorithm 1 is denoted as R--SIDCO.

Ideally, to create optimally incoherent unit norm frames $\mathbf{F}$ of size $m \times N$ we would like to solve exactly the following optimization problem:
		\begin{equation}
			\underset{\mathbf{f}_i; \ \| \mathbf{f}_i \|_2 = 1, \ 1 \leq i \leq N}{\text{minimize}} \ \  \underset{i,j;\ j \neq i}{\max \ }{|\mathbf{f}_j^T \mathbf{f}_i|.}
			\label{eq:IDCOMain}
		\end{equation}
The difficulty lies in the fact that both the objective and the equality constraint are non-convex. The approach taken in \cite{OurPaper} is not to solve this problem directly, but given a real unit norm frame $\mathbf{H} = \begin{bmatrix} \mathbf{h}_1 & \dots & \mathbf{h}_N \end{bmatrix} \in \mathbb{R}^{m \times N}$ to find a new real unit norm frame $\mathbf{F} = \begin{bmatrix} \mathbf{f}_1 & \dots & \mathbf{f}_N \end{bmatrix}$ of equal size, near the initial one, with smaller mutual coherence. Therefore, we update each frame vector $\mathbf{h}_i$ at a time to a new vector $\mathbf{f}_i$ such that its maximum absolute value inner product with all the other frame vectors $\mathbf{h}_j, j \neq i$ is minimized. The problem for each $\mathbf{h}_i$ is
		\begin{equation}
			\underset{\mathbf{f}_i \in \mathbb{R}^{m}; \ \|\mathbf{f}_i - \mathbf{h}_i\|_2^2 \leq T_i}{\text{minimize}} \ \  \underset{j;\ j \neq i}{\max} \ | \mathbf{h}_j^T\mathbf{f}_i |.
			\label{eq:IDCOMain2}
		\end{equation}
We normalize and update $\mathbf{h}_i = \mathbf{f}_i \| \mathbf{f}_i \|_2^{-1}$ after solving \eqref{eq:IDCOMain2}. Also, we can define 
\begin{equation}
	\mathbf{H}_i = \begin{bmatrix} \mathbf{h}_1 & \dots & \mathbf{h}_{i-1} & \mathbf{h}_{i+1} & \dots & \mathbf{h}_N \end{bmatrix} \in \mathbb{C}^{m \times (N-1)},
	\label{eq:theH}
\end{equation}
for each $i = 1,\dots,N$ and thus the objective function becomes $\| \mathbf{H}_i^T \mathbf{f}_i \|_\infty$. To solve this problem, we define for each vector $\mathbf{h}_i$ in the reference frame a trust region (an $m$-ball centered at $\mathbf{h}_i$ of radius $\sqrt{T_i}$), where we search for a new vector $\mathbf{f}_i$ such that its correlation with the other vectors in the frame $\mathbf{H}_i$ is smaller than that of $\mathbf{h}_i$. Since now we are dealing with a convex objective function and constraint, the formulation in \eqref{eq:IDCOMain2} can be viewed as a convex relaxation of the original problem \eqref{eq:IDCOMain}. The parameters $T_i$ are chosen such that
\begin{equation}
	T_i \leq 1 - \underset{j;\ j \neq i}{\max}\ |g_{ij}|^2, \text{ with } g_{ij} \text{ the entries of } \mathbf{G} = \mathbf{H}^T \mathbf{H}.
	\label{eq:chooseT}
\end{equation}
This inequality establishes the maximum possible value of $T_i$ such that the current variable $\mathbf{f}_i$ (scaled by any constant $\alpha \neq 0$) remains in an $m$-ball around $\mathbf{h}_i$ constrained by the vectors of $\mathbf{H}_i$. This constraint guarantees a monotonically decreasing mutual coherence, and as numerically shown in \cite{OurPaper} the convergence is fast, in a few steps. Since we observe, in numerical simulations, that local minima are found quickly, the heuristic Step 2 in Algorithm 1 updates the frames such that there might be (and experimentally we observe that it usually is) a temporary increase in the mutual coherence, but the frame is closer, in Frobenius norm, to a tight frame. The temporary increase in coherence is attenuated in the following, regular, steps of the algorithm.

Based on the general template presented here, we formulate and solve similar optimization problems where the entries satisfy additional constraints in subsequent sections.

\begin{algorithm}[t]
\caption{ \textbf{-- Sequential Iterative Decorrelation by Convex Optimization (R--SIDCO) \cite{OurPaper} } \newline \textbf{Input: } The pair $(m, N)$ and the number of iterations $K$. \newline \textbf{Output: } Frame $\mathbf{H} \in \mathbb{R}^{m \times N}$ as incoherent as possible until maximum number of iterations $K$ is reached.}
\begin{algorithmic}

\State \textbf{Initialization:}
\begin{enumerate}
	\setlength{\itemindent}{+.1in}
	\item Create $\mathbf{H} \in \mathbb{R}^{m \times N}$ with random entries from the standard Gaussian distribution. Normalize its columns.

	\item With $\mathbf{H} = \mathbf{U \Sigma V}^H$ update the frame by the unit polar decomposition (by Theorem 2 of \cite{AlterningProjection}): $\mathbf{H} = \mathbf{U}\mathbf{V}^H$. Normalize its columns.
\end{enumerate}

\State \textbf{Iterations }$1,\dots ,K$:
	\begin{enumerate}
	\setlength{\itemindent}{+.1in}

		\item For each $i$ in randomized $\{ 1,\dots,N \}$:
		\begin{enumerate}
			\item Set radius $T_i$ of trust region to the bound in \eqref{eq:chooseT} and solve \eqref{eq:IDCOMain2} for $\mathbf{f}_i$.

			\item Normalize and update $\mathbf{h}_i = \mathbf{f}_i \| \mathbf{f}_i \|_2^{-1}$.

		\end{enumerate}

		\item If the algorithm has converged, update frame by its unit polar decomposition $\mathbf{H} = \mathbf{UV}^H$ (Theorem 2 of \cite{AlterningProjection}) and normalize columns.

	\end{enumerate}

\end{algorithmic}
\end{algorithm}

\section{Extensions of R--SIDCO}

This section is concerned with extending the R--SIDCO method proposed for real valued incoherent frame design to other frame types: complex valued, complex unital, nonnegative both real and complex and sparse both real and complex. All the extensions we discuss focus on some modifications of the template Algorithm 1 and we deal with each one separately in the next sections.

\subsection{Complex valued incoherent frames}

The easiest extension of R--SIDCO is to complex valued frames $\mathbf{H} \in \mathbb{C}^{m \times N}$. Consider the optimization problem
		\begin{equation}
			\underset{\mathbf{f}_i  \in \mathbb{C}^{m}; \ \|\mathbf{f}_i - \mathbf{h}_i\|_2^2 \leq T_i}{\text{minimize}} \ \   \|\mathbf{H}_i^H\mathbf{f}_i \|_\infty.
			\label{eq:C-IDCOMain2}
		\end{equation}
In this complex case, the optimization problem is a quadratic program with $2m+1$ real variables and $N-1$ constraints. With the choice of $T_i$ from the bound \eqref{eq:chooseT}, each update of $\mathbf{f}_i$ is guaranteed to keep or decrease the mutual coherence of the frame (see Remark 1 of \cite{OurPaper}). Briefly, this is because the trust region constraint leads to a higher $\ell_2$ norm of $\mathbf{f}_i$ with increasing the angle from the reference $\mathbf{h}_i$ and therefore the normalization step $\mathbf{f}_i \| \mathbf{f}_i \|_2^{-1}$ cannot increase the mutual coherence. As such, the convergence to a local minimum is guaranteed. We call this approach C--SIDCO.

Frames designed by C--SIDCO find use in communication applications since they are equivalent to antipodal spherical codes \cite{BCASC}. Also, in the quantum information theory literature there are constructions of symmetric, informationally complete, positive operator valued measures (SIC-POVM) \cite{SICPOVM} that achieve coherence $1/\sqrt{m+1}$ for $N = m^2$ vectors in complex Hilbert spaces. These complex frames have been constructed, by numerical methods, for all dimensions $m\leq 151$ (and a few others up to $m = 844$) \cite{LatestSIC} and it is conjectured that  they exist for any $m$ (Zauner's conjecture). Notice that in this general case, SIC-POVMs supply highly overcomplete frames with very low coherence.

\subsection{Complex unital incoherent frames}

An additional constraint that all entries have equal magnitude can be added to the design of complex valued incoherent frames. Together with the constraint $\| \mathbf{f}_i \|_2 = 1$ we have that $| f_{ij} | = m^{-1/2}$, which is also a non-convex constraint. Thus, after relaxing the magnitude constraints, the proposed problem is
		\begin{equation}
			\underset{\mathbf{f}_i  \in \mathbb{C}^{m};\ | f_{ij} - h_{ij} |^2 \leq T_i,\ |f_{ij} | - m^{-1/2} \leq  \gamma}{\text{minimize}} \ \  \|\mathbf{H}_i^H\mathbf{f}_i \|_\infty, \\
			\label{eq:U-IDCOMain2}
		\end{equation}
with the constraints for all $j = 1,\dots,m$ and where $0< \gamma \ll 1$ is a constant that manages the unital constraint. In this formulation, we focus on the individual entries of the variable $\mathbf{f}_i$ since their magnitudes are also constrained. The new constraint $| f_{ij} - h_{ij} |^2 \leq T_i$ implicitly imposes $\| \mathbf{f}_i - \mathbf{h}_i \|_2^2 \leq m T_i$. In the spirit of the trust regions approach, we approximate the constraint $|f_{ij}| = 1$ by defining a trust region with the convex constraint $|f_{ij} | - m^{-1/2} \leq  \gamma$. The constraints in \eqref{eq:U-IDCOMain2} ensure together that the trust region is defined around the reference vector and close to unit magnitude entries. After solving this problem the normalization step is $\mathbf{f}_i  =m^{-1/2}\left( \mathbf{f}_i \oslash |\mathbf{f}_i| \right)$, where $\oslash$ is the elementwise division operation. Unfortunately, due to the new trust regions the convergence result previously presented for C--SIDCO does not hold anymore. Therefore, in the iterative process we keep track of the best frame achieved so far (the one with the lowest mutual coherence) and return it when the algorithm terminates.

In the case of complex unital frames, the decomposition that takes place at Step 2 of the initialization in C--SIDCO destroys in general the unital structure previously imposed. Thus, Step 2 is changed to a decomposition $\mathbf{H} = \mathbf{UV}^H$, with $\mathbf{H} = \mathbf{U\Sigma V}^H$, and a normalization $\mathbf{H} = m^{-1/2}( \mathbf{H} \oslash |\mathbf{H}|)$.

We do expect these complex unital frames to achieve coherence larger than the general complex frames due to the additional unital constraint. As such, the complex valued incoherent frames, after normalization, serve as good initializations in these situations. Alternatively, we can also use incoherent harmonic frames, which are discussed next, since they are also naturally unital.

This approach is denoted as U--SIDCO, and it produces frames that find use in communication systems, where a low \textit{peak-to-average-power} ratio (PAPR) is desirable, for example in limited feedback codebooks \cite{CodeBooks}. In practice, constant amplitude signals are also used by power-limited hardware \cite{AlterningProjection}. In the frame literature there is a work on dealing with a low PAPR \cite{AlterningProjection}. Our algorithm U--SIDCO designs incoherent frames with $\text{PAPR} =1$, the lowest value. 

\subsection{Nonnegative incoherent frames}

Another extension of R--SIDCO is to nonnegative frames, i.e., frames with nonnegative entries, both real and complex (separately on each component). The optimization problem to be solved is:
		\begin{equation}
			\underset{\mathbf{f}_i; \ \|\mathbf{f}_i - \mathbf{h}_i\|_2^2 \leq T_i,\ \Re(\mathbf{f}_i) \geq \mathbf{0},\ \Im(\mathbf{f}_i) \geq \mathbf{0}}{\text{minimize}} \ \   \|\mathbf{H}_i^H\mathbf{f}_i \|_\infty.
			\label{eq:PC-IDCOMain2}
		\end{equation}
The convergence (to a local minimum) results from the general C--SIDCO and R--SIDCO hold in this case since the nature of the trust region around the references $\mathbf{h}_i$ is not modified by the additional nonnegativity constraints and there is no post processing of the solutions $\mathbf{f}_i$. The difficulty lies in applying the heuristic Step 2, since there are no guarantees that the unit polar decomposition $\mathbf{UV}^H$ leads to a nonnegative frame. This is because  the original frame $\mathbf{H}$ was nonnegative -- indeed experimentally it is observed that this does not usually happen.  To circumvent this difficulty we propose a new simple update: $\mathbf{H} = \mathbf{H} + \delta \mathbf{R}$, where $\mathbf{R}$ of size $m \times N$ is a matrix with random entries drawn from a standard Gaussian distribution; and normalize the columns of $\mathbf{H}$ to have unit $\ell_2$ norm. Intuitively, we add to the current reference frame $\mathbf{H}$ a random small perturbation -- $\delta$ controls the size of the perturbation. We consider that $|\delta| \ll 1$ and we are that it is not possible to reach negative entries close to zero, since the following optimization problems impose again the nonnegativity constraint.

These approaches are denoted by NR--SIDCO and NC--SIDCO,  in the real and complex cases, respectively. Nonnegative frames are useful for nonnegative matrix factorizations \cite{SNMF} applications.

\subsection{Sparse incoherent frames}

We now consider sparse frames, which contain a large number of zero entries, which are also incoherent. We propose to solve the optimization problem:
		\begin{equation}
			\underset{\mathbf{f}_i; \ \|\mathbf{f}_i - \mathbf{h}_i\|_2^2 \leq T_i}{\text{minimize}} \ \   \|\mathbf{H}_i^H\mathbf{f}_i \|_\infty + \lambda \| \mathbf{f}_i \|_1.
			\label{eq:SC-IDCOMain2}
		\end{equation}
The parameter $\lambda$, which is fixed and provided as an input, controls the $\ell_1$ regularization term which introduces the zero entries in the current variable frame vector $\mathbf{f}_i$, in the style of the LASSO \cite{tibshirani1996regression}. Various $\lambda$ produce frames of differing sparsity levels. 

When the number of iterations $K$ has been reached or the algorithm has converged, a polishing step follows. For each $\mathbf{h}_i$, we establish its support $\mathcal{S}_i = \{ k \ | \ |h_{ki} | \leq \epsilon \text{ with } k = 1,\dots,m \}$, and then solve the problem
		\begin{equation}
			\underset{\mathbf{f}_i;\ \|\mathbf{f}_i - \mathbf{h}_i\|_2^2 \leq T_i, \ f_{ki} = 0 \ \forall k \in \mathcal{S}_i}{\text{minimize}} \ \   \|\mathbf{H}_i^H\mathbf{f}_i \|_\infty.\\
			\label{eq:SC-IDCOMain2b}
		\end{equation}
When designing sparse incoherent frames, the heuristic Step 2 of Algorithm 1 is avoided, since it will not preserve the sparsity. We have observed numerically that a very good initialization is a general frame created by R--SIDCO or C--SIDCO (equivalent to having $\lambda = 0$ in \eqref{eq:SC-IDCOMain2}). These new approaches are denoted by SR--SIDCO and SC--SIDCO, respectively.

An alternative to the $\ell_1$ approach in \eqref{eq:SC-IDCOMain2} is to decide a priori the zero entries of each frame vector and run the optimization problems only for the other entries, in the same manner described in \eqref{eq:SC-IDCOMain2b}. Let us consider for example the case where we impose the zero structure to the frame $\mathbf{H} \in \mathbb{C}^{m \times N}$ as follows:
$	\mathbf{H} = \begin{bmatrix}
				\mathbf{H}_{11} & \mathbf{0} \\
				\mathbf{0} & \mathbf{H}_{22}
				\end{bmatrix}$,
where each block has size $m/2 \times N/2$, assuming for simplicity no rounding issues. We have that $\mu(\mathbf{H}) = \max\{ \mu(\mathbf{H}_{11}), \mu(\mathbf{H}_{22}) \}$, and also that the Welch bound for frames $(m/2, N/2)$ is $\sqrt{2 \frac{N-1}{N-2}}$ larger than the Welch bound for frames $(m, N)$. As such, no matter how the optimization is done for the entries of $\mathbf{H}_{11}$ and $\mathbf{H}_{22}$, there is a lower bound to the minimum for the overall frame. Also, the structure imposed can lead to low values for spark$(\mathbf{H})$ (the minimum number of linear dependent columns of $\mathbf{H}$) whose value should be as large as possible to guarantee the success of sparse recovery algorithms \cite{Spark}. A simple idea is to spread the zeros in the frame coefficients, for each vector for example, in a random manner. We observe, by numerical experimentation, that heuristics like these help avoid the problems previously described. Additionally, this idea of explicitly setting certain entries to zero for each frame vector has the advantage of providing control on the sparsity level of each frame vector individually. This way, we can ensure that all frame vectors have the same sparsity level, a task which is difficult when using the $\ell_1$ penalty. A final advantage for fixing the zero entries a priori is that the convergence results of R--SIDCO and C--SIDCO hold (they hold also for the polishing steps previously described).

We can find in the frame literature work on sparse tight frames created by spectral tetris \cite{SpectralTetris} with less than $3N$ nonzero entries, sparse Steiner equiangular tight frames \cite{Steiner} with less than $\sqrt{2m}N$ nonzero entries, and an approach based on discrete Gabor expansions \cite{Strohmer1998}. The goal of these constructions is to reduce the computational burden of using the frame as an analysis/synthesis operator.

We develop next a different numerical approach for the design of incoherent frames that are selections from fixed unital matrices.

\section{Harmonic frames}

A harmonic frame $\mathbf{H} \in \mathbb{C}^{m \times N}$ is a frame consisting of a subset of $m$ rows from the Fourier matrix $\mathbf{F} \in \mathbb{C}^{N \times N}$. Due to their structural properties, we propose an algorithm to build incoherent harmonic frames that is significantly different from the SIDCO approaches.

The Gram matrix of any harmonic frame is circulant, that is, 
$	\mathbf{G} =  \mathbf{H}^H \mathbf{H} = \text{circ}(\mathbf{g}) = m^{-1}\mathbf{F}^H \text{diag}(\mathbf{\tilde{g}}) \mathbf{F}$,
where $\mathbf{g} = m^{-1} \mathbf{F}^H \mathbf{\tilde{g}}$ is the first column of $\mathbf{G}$, and $\mathbf{\tilde{g}} \in \{0,1\}^N$ denotes a binary vector corresponding to selecting certain rows of the Fourier matrix, i.e., $\mathbf{H} = \text{diag}(\mathbf{\tilde{g}})\mathbf{F}$. The circulant matrix $\mathbf{G}$ is completely defined by its first column vector $\mathbf{g}$ -- all other columns are cyclic permutations of this vector, with offset equal to the column index. Given a set $\mathcal{K} \subset \mathbb{Z}_N$, we call a selection pattern a vector $\mathbf{\tilde{g}} \in \{0,1\}^N$ such that $\tilde{g}_{k+1} = 1$ when $k \in \mathcal{K}$ and zero otherwise. We equivalently denote $\mathbf{\tilde{g}} = \mathds{1}_\mathcal{K}$. This vector/set equivalence notation will ease the presentation of the results in this section.

There is a strong connection between harmonic ETFs and difference sets \cite{DifferenceSets} (a $(N, m, \lambda)$-difference set is a subset of size $m$ of a larger set of size $N$ such that every nonzero element of this set can be represented as a difference between two elements in exactly $\lambda$ ways). Other work uses character sums estimates \cite{CSWithFourier} or almost difference sets \cite{ADS} to construct highly incoherent harmonic frames for certain dimensions. Unfortunately, these approaches cover only some values of $(m,N)$. We now propose an optimization procedure that allows for the design of highly incoherent harmonic frames of any dimension.

\subsection{Incoherent harmonic frames}

For any $(m, N)$, to construct the most incoherent harmonic frame, we would like to solve exactly the non-convex optimization problem:
		\begin{equation}
			\underset{\mathbf{\tilde{g}};\ \sum_{k=1}^N \tilde{g}_k = m,\ \tilde{g}_k \in \{0,1\}}{\text{minimize}} \ \ m^{-1} \| \mathds{F}\mathbf{\tilde{g}} \|_\infty,
			\label{eq:HarmonicETF}
		\end{equation}
where the matrix $\mathds{F}$ is the complex conjugate transpose Fourier matrix $\mathbf{F}^H$ of size $N$ restricted to the rows index by $2:\lfloor N/2 \rfloor+1$. The objective function is the mutual coherence of the harmonic frame created by selecting the rows of the Fourier matrix corresponding to ones in the vector $\mathbf{\tilde{g}}$. The absolute value Gram matrix of a harmonic frame is a symmetric circulant matrix; subsequently, the search for its maximum off-diagonal entry, i.e., the coherence, is restricted to the first half (except the first entry, which is 1) of the first column. This simplification is reflected in the structure of the operator $\mathds{F}$. Solving exactly the non-convex (due to the integer constraints) optimization problem in \eqref{eq:HarmonicETF}, will result in deciding which $m$ rows of the Fourier matrix deliver the smallest mutual coherence. In this setting, $m$ is given (fixed) and the problem can be seen as a mixed binary optimization problem.

The binary problem in \eqref{eq:HarmonicETF} is hard because the search space is large, combinatorial. The complexity can be reduced since there are several equivalence classes. If we select rows of the Fourier matrix from a set $\mathcal{K} \subset \{1,\dots,N\}$ of size $| \mathcal{K} | = m$, with associated selection pattern $\mathbf{\tilde{g}} = \mathds{1}_\mathcal{K}$, we have a particular coherence value. This value is also reached when: the set is circularly shifted $\mathcal{K}_\text{new} = (\mathcal{K} + j) \mod N$ for $j=1,\dots,N-1$ (the result follows from the circular shift property of the Fourier transform), and the set is multiplied with a constant $\tau \in \mathbb{Z}_N$, relatively prime to $N$, i.e, $\mathcal{K}_\text{new} = \tau \mathcal{K} \mod N$. Proofs of these facts are given in \cite[Theorem 4]{FullSparkFrames}. This shows that the solution to \eqref{eq:HarmonicETF} is not unique. Therefore we let $\tilde{g}_1 = 1$ be fixed when solving \eqref{eq:HarmonicETF}.

Furthermore, a result on the set $\mathcal{K}$ offers insights into the result of the complementary set $\mathcal{K}_\text{c}  = \{ 0, 1, \dots,N-1 \} \setminus \mathcal{K}$. Consider the new selection pattern $\mathbf{\tilde{g}}_\text{c} = \mathds{1}_{\mathcal{K}_\text{c}} = \mathbf{1}  - \mathbf{\tilde{g}}$. The unique part of the circulant Gram matrix of the frame resulting from the new selection pattern is $(m-N)^{-1}\mathbf{F}^H \mathbf{\tilde{g}}_\text{c} 
							       = (m-N)^{-1} \begin{bmatrix} m-N \\ m\mathbf{g}_{2:N} \end{bmatrix} = \begin{bmatrix} 1 \\ \frac{m}{m-N} \mathbf{g}_{2:N} \end{bmatrix}.$
This result establishes the fact that finding an incoherent (or ETF) harmonic frame $(m,N)$ automatically leads to the construction of an incoherent (or ETF) harmonic frame $(N-m, N)$.

\subsection{An algorithm for the design of incoherent harmonic frames}

Based on \eqref{eq:HarmonicETF}, consider the following convex optimization problem:
\begin{equation}
	\begin{aligned}
	& \underset{\mathbf{\tilde{g}}}{\text{minimize}} & & m^{-1} \| \mathds{F}\mathbf{\tilde{g}} \|_\infty + \lambda \|\mathbf{W\tilde{g}}  \|_1 \\
	& \text{subject to} & & \tilde{g}_1  = 1, \ \sum_{k=2}^N \tilde{g}_k = m-1,\ \tilde{g}_k = 0, \text{ for } k \in \mathcal{K}_0 \\
	& & & 0 \leq \tilde{g}_k \leq 1, \text{ for } 2 \leq k \leq N.
	\label{eq:ConvexHarmonicETF}
	\end{aligned}
\end{equation}
This optimization problem is at the heart of the proposed optimization procedure based on iterative reweighted $\ell_1$ optimization (IRL1) \cite{IRL1} and solved via CVX \cite{CVX}. The overall proposed design algorithm is shown in Algorithm 2.

Compared to \eqref{eq:HarmonicETF}, we add a regularization term $\|\mathbf{W\tilde{g}}\|_1$ to promote sparsity for the solution $\mathbf{\tilde{g}}$. The diagonal matrix $\mathbf{W}$ is fixed and consists of weights inverse proportional to the magnitudes of the entries in $\mathbf{\tilde{g}}$, used to further increase the sparsity of the solution. This regularization term works in connection with the sum constraints. The overall goal of the regularization term is to move the solution's coefficients either towards zero or one.

Regarding the constraints, the key observation is that we relax the hard binary constraints $\tilde{g}_k \in \{0, 1\}$ to the convex inequality constraints $\tilde{g}_k \in [0,1]$. Notice that due to this relaxation, the solution $\mathbf{\tilde{g}}$ may not have exactly $m$ nonzero entries but possibly more. The last constraint explicitly imposes zeros in the position indexed by a given set $\mathcal{K}_0$. Without this constraint, in the first step of Algorithm 2 with $\mathbf{W} = \mathbf{I}$, the solution to \eqref{eq:ConvexHarmonicETF} would always be $\mathbf{\tilde{g}} = \begin{bmatrix} 1; & \frac{m-1}{N-1}\mathbf{1} \end{bmatrix}$. To see this start from
\begin{equation*}
	m^{-1} \mathbf{F}^H \begin{bmatrix}
				1 \\ \frac{m-1}{N-1}\mathbf{1}
				\end{bmatrix}  =  m^{-1} \begin{bmatrix}
							m \\ \left( 1 - \frac{m-1}{N-1}\mathbf{1} \right) 
							\end{bmatrix} = \begin{bmatrix}
										1 \\ \frac{N-m}{m(N-1)}\mathbf{1}
										\end{bmatrix},
\end{equation*}
and observe that $\frac{N-m}{m(N-1)} \leq \sqrt{\frac{N-m}{m(N-1)}}$ for any $(m, N)$ with $m \leq N$. Thus, for this $\mathbf{\tilde{g}}$ the $\ell_\infty$ norm part of the objective function is always below the Welch bound. In fact, this solution provides the lowest infinity norm, and thus, it is preferred to any other solution. Because of this behavior, without the constraint with $\mathcal{K}_0$, the optimization problem would not introduce zeros in any position. Therefore, as a heuristic measure, we explicitly impose zeros in the solution by the set $\mathcal{K}_0$, whose size we choose to be $\lceil \zeta(N-m) \rceil$, with $\zeta \ll 1$ fixed.

The key idea of Algorithm 2 is to set a small number of coefficients to zero and let the optimization problem decide on the others to be nulled, such that the coherence is minimized. Since different sets $\mathcal{K}_0$ produce different results we run this optimization procedure with several set choices.

\begin{algorithm}[t]
\caption{ \textbf{-- IRL1 Incoherent Harmonic Design. } \newline \textbf{Input: } The pair $(m, N)$, the number of iterations $K$, parameters $\zeta$ and $\lambda$ and the length of the local search $\ell$. \newline \textbf{Output: }  The binary vector $\mathbf{\tilde{g}}$ such that the frame $\text{diag}(\mathbf{\tilde{g}})\mathbf{F}$ has mutual coherence as low as possible and $\mathbf{1}^T \mathbf{\tilde{g}} = m$.}
\begin{algorithmic}
	\State \textbf{1. } Generate randomly the set $\mathcal{K}_0 \subset \{2,\dots, N\}$ of size $\left\lceil \zeta (N-m) \right\rceil$.

	\State \textbf{2. } Set $\mathbf{W} = \mathbf{I}$.

	\State \textbf{3. } For $1,\dots, K$: solve \eqref{eq:ConvexHarmonicETF} for fixed $\mathbf{W}$ and $\mathcal{K}_0$, update diagonal $w_k = 1 - \tilde{g}_k$.

	\State \textbf{4. } Establish the support of $\mathbf{\tilde{g}}$: $\mathcal{K} = \{ k-1 \ | \ | \tilde{g}_k | > \epsilon \text{ with } k=1,\dots,N \}.$

	\State \textbf{5. } If necessary, reduce the support size $|\mathcal{K}|$ to $m$:
				\begin{equation}
					\begin{aligned}
						& \quad \text{for } | \mathcal{K} | \text{ down to } m \text{ set } \mathcal{K} = \mathcal{K} \setminus \{k^* \} \text{ with } \\
						& \quad  \quad  \quad  \quad  \quad  \quad k^* = \underset{\mathcal{K}' = \mathcal{K} \setminus \{ k \} \text{ for each } k \in \mathcal{K}}{\text{arg min}} \| \mathds{F} \mathds{1}_{\mathcal{K}'} \|_\infty .
					\end{aligned}
				\end{equation}

	\State \textbf{6. } Start a local search close to the set $\mathcal{K}$:
				\begin{equation}
					\begin{aligned}
						\{ \mathcal{A}, \mathcal{Z} \} = \underset{\mathcal{A} \subset \{ \mathcal{K}_\text{c} \cup \mathcal{Z}\}, |\mathcal{A}| = \ell}{\text{arg min}} \| \mathds{F} \mathds{1}_{\mathcal{K}  \setminus \mathcal{Z} \cup \mathcal{A}  }\|_\infty,  \\ \text{ for each set } \mathcal{Z} = \{ z_1, \dots, z_\ell \} \subset \mathcal{K}, |\mathcal{Z}| = \ell.
					\end{aligned}
				\end{equation}

	\State \textbf{7. } Set $\mathcal{K} = \mathcal{K} \setminus \mathcal{Z} \cup \mathcal{A}$ and return $\mathbf{\tilde{g}} = \mathds{1}_\mathcal{K}$.
\end{algorithmic}
\end{algorithm}
Notice that due to the constraints on the variable $\mathbf{\tilde{g}}$, the $\ell_1$ penalization term simplifies to
	$\|\mathbf{W\tilde{g}}\|_1 = \mathbf{w}^T\mathbf{\tilde{g}}$,
where $\mathbf{w} = \text{diag}(\mathbf{W})$. A discussion about the weights is in order. Taking absolute values is omitted due to the positive entries in the solution. In this implementation we use the update $w_k = (\tilde{g}_k + \epsilon)^{-1}$, for  $1\leq k \leq N,$ where $\epsilon$ is a fixed constant close to zero. The idea of the weights is to have magnitudes inverse proportional to the magnitudes of the entries in the solution. Analysis of the IRL1 is difficult in general, but some results are available \cite{IRL1Analysis}. 
Other ways of choosing the weights avoid the use of an additional parameter. For example, consider the update $w_k = 1 - \tilde{g}_k \| \mathbf{\tilde{g}} \|_\infty^{-1},$ for $1\leq k \leq N$. In this case, the penalty term reaches a stationary point of
\begin{equation}
	\mathbf{w}^T \mathbf{\tilde{g}} = \| \mathbf{\tilde{g}} \|_1 - \frac{\| \mathbf{\tilde{g}} \|_2^2  }{ \| \mathbf{\tilde{g}} \|_\infty} = m - \| \mathbf{\tilde{g}} \|_2^2,
\end{equation}
where the last equality holds in our case due to the constraints on $\mathbf{\tilde{g}}$. If the entries of the solution are binary, i.e., in $\{0,1\}$, then the penalty term reaches its minimum and is exactly zero. Notice that the term is concave, and thus it has been shown that its minimization is NP--hard in general \cite{BinaryIsNPHard}. The iterative steps proposed here approximate the problem by solving a sequence of convex optimization problems. This way we make the overall problem tractable, without the guarantee of reaching the global optimum solution.

The convex optimization problem \eqref{eq:ConvexHarmonicETF} is used iteratively in Algorithm 2. The first step is to randomly generate the set $\mathcal{K}_0$ with the indices in the solution that are explicitly set to zero. As explained before, this is to avoid the optimization problem reaching a trivial solution. Next, the reweighted $\ell_1$ optimization follows for $K$ iterations trying to pushing the solution's coefficients to zero or one values.

Since IRL1 does not control directly the support of the solution, Step 5 iteratively eliminates, i.e., sets to zero, coefficients from $\mathbf{\tilde{g}}$ (or equivalently the set $\mathcal{K}$), until only $m$ remain. With this strategy, each elimination causes the minimal increase in the mutual coherence. Step 6 makes a last effort to further decrease the mutual coherence. Based on the solution computed up to this point $\mathds{1}_\mathcal{K}$, we search for a better solution ``around" this reference. The search, combinatorial in nature, checks for the best subset $\mathcal{Z}$ of size $\ell$ of $\mathcal{K}$ that can be substituted by a new set $\mathcal{A}$ of equal size, such that the mutual coherence is maximally reduced. For computational reasons, we choose $\ell \ll m$. Assuming that $\mathbb{F} \mathds{1}_\mathcal{K}$ has been computed and $\mathcal{K}' = \{\mathcal{K} \setminus \{ j \} \} \cup \{ k \} $, observe that $\mathbb{F} \mathds{1}_{\mathcal{K}'} = \mathbb{F} \mathds{1}_\mathcal{K} - \mathbb{F} \mathbf{e}_j + \mathbb{F} \mathbf{e}_k$. This shows that the local search procedure in Step 6 of Algorithm 2 can be done fast by updating the unique part of the Gram matrix across iterations, without fully reconstructing it at each step.

The proposed method can be extended to the selection of rows of any fixed matrix, provided that its entries have constant magnitude, for example Hadamard matrices. Otherwise, the normalization operation in \eqref{eq:MutualCoherence1} is necessary and a different optimization strategy needs to be proposed.

\section{Numerical results}

We provide numerical results for the proposed algorithms, comparing with the Welch bound and with other previously proposed algorithms. We also show how the created incoherent frames perform when used to recover sparse vectors.

\subsection{Incoherent frames designed via the proposed methods}

Equiangular tight frames exist only for a small number of pairs $(m, N)$. For example, in the case of harmonic frames for $N\leq 256$, there are 143 known ETFs (checked by the existence of the equivalent difference set \cite{Jolla}). Even for the pairs where we encounter an ETF, usually the redundancy is quite low. For example, out of the 143 known harmonic ETFs, 57 have redundancy $\rho \in [2, 3]$, while harmonic ETFs with high redundancy ($\rho\geq 10$) exist only for a few $m$. In general, harmonic ETFs mostly exist for a prime or prime power $m$. A similar discussion can be made in the case of real valued ETFs. Conditions developed in \cite{ETF2} for the existence of real valued ETFs allow for 182 of such structures when $N \leq 256$. Out of these, 97 ETFs have redundancy $\rho \in [2,3]$. A recent survey of ETFs can be found in \cite{LatestTable}. We now present numerical results for constructing highly incoherent frames (not necessarily equiangular) for any pair $(m, N)$.
\begin{figure}[t]
	\centering
	\begin{subfigure}{0.42\textwidth}
		\includegraphics[trim = 5 0 30 20, clip, width=\textwidth]{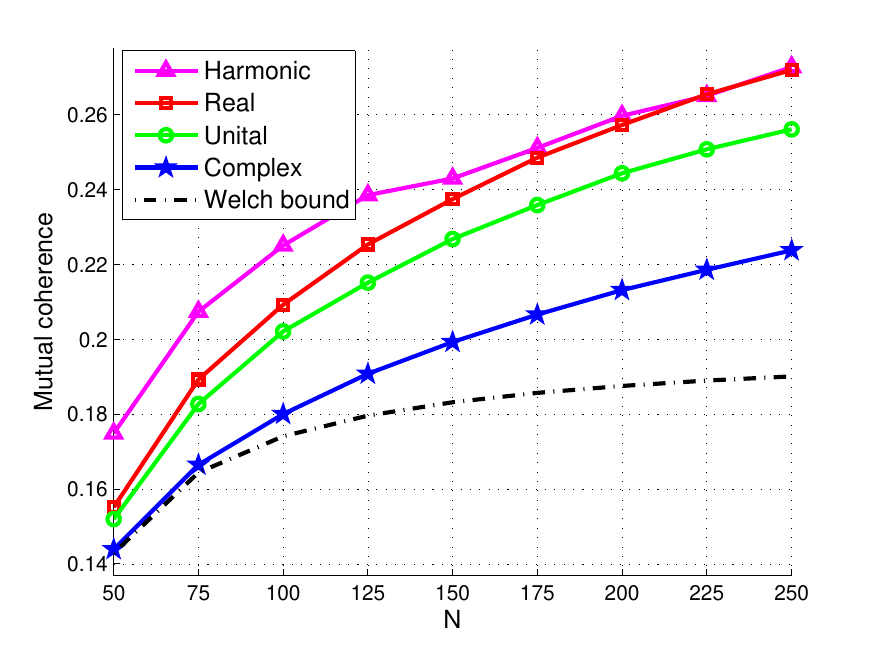}
	\end{subfigure}
	\centering
	\begin{subfigure}{0.42\textwidth}
		\includegraphics[trim = 5 0 30 20, clip, width=\textwidth]{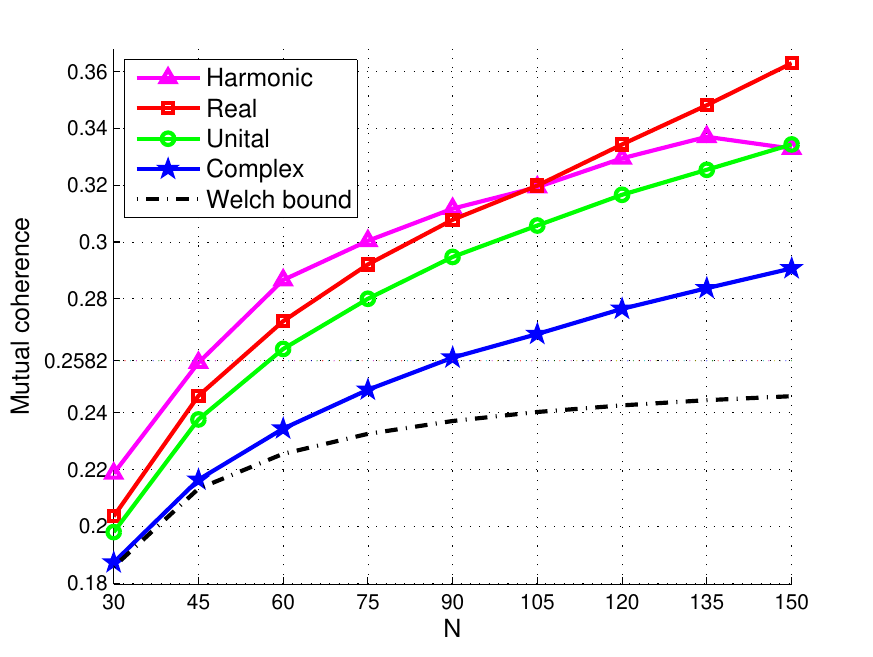}
	\end{subfigure}
	\caption{Incoherent frames designed for $m = 25$ (left) and $m = 15$ (right).}
	\label{fig:frames25and15}
\end{figure}
\begin{figure}[t]
\centering
\begin{subfigure}{0.42\textwidth}
\includegraphics[trim = 5 0 30 20, clip, width=\textwidth]{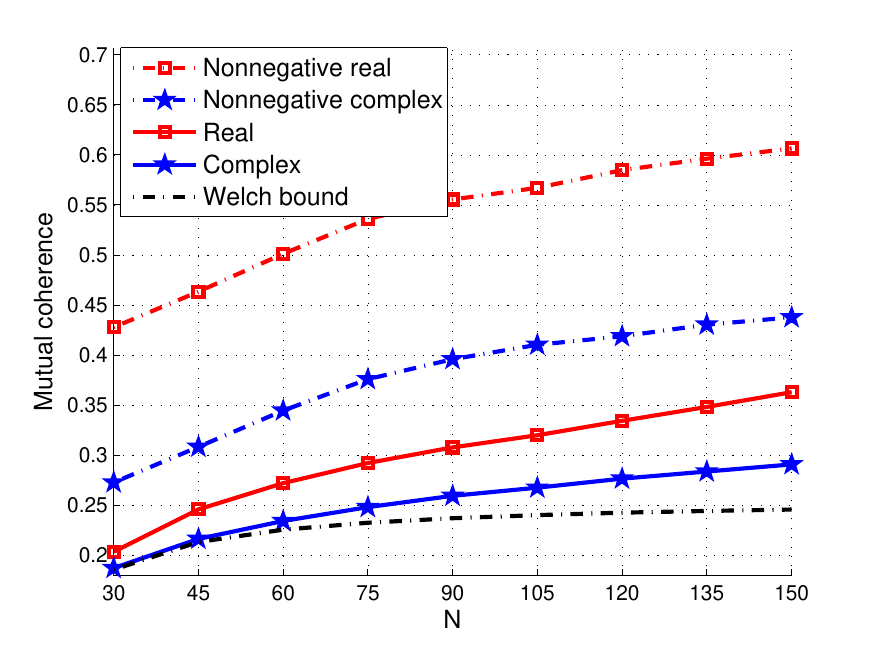}
\label{fig:framessparse15}
\end{subfigure}
\centering
\begin{subfigure}{0.42\textwidth}
\includegraphics[trim = 5 0 30 20, clip, width=\textwidth]{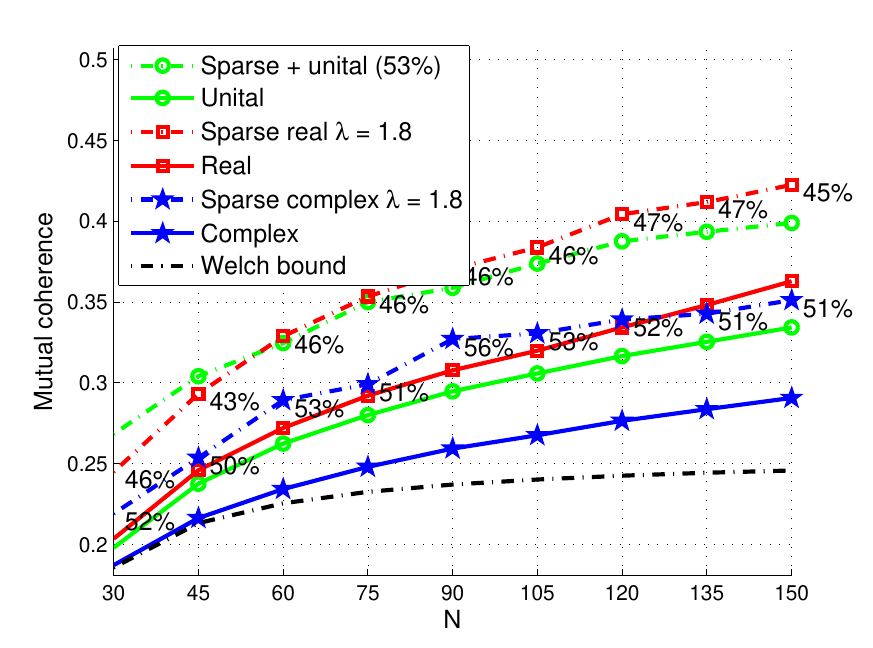}
\label{fig:framespositive15}
\end{subfigure}
\caption{Incoherent frames designed for $m=15$.}
\label{fig:frames15}
\end{figure}

Figure \ref{fig:frames25and15} shows the mutual coherence for various frame types and various $N$ with fixed $m = 25$ and $m = 15$ (for the latter we also show sparse and positive frames in Figure \ref{fig:frames15}). The Welch bound is also shown as a reference. In terms of the incoherent frames, the best results are reached by the general complex frames designed via C--SIDCO. For redundancies $\rho \in \{2,3,4\}$, the results are close to the Welch bound, while for larger values the gap to the bound increases. For low redundancies, the complex frames designed via C--SIDCO show coherence levels below the $1/\sqrt{m}$ limits of 0.2 and 0.2582 (for $m=25$ and $m = 15$ respectively), that are marked in the plots. The next best frames in terms of coherence are the complex unital ones designed via U--SIDCO with $\gamma = 0.01$. The performance gap between them and the similar harmonic incoherent frames seems to decrease with $N$. Of course, complex unital frames must always produce structures at least as incoherent as the harmonic frames. Failure to do so is attributed entirely to the numerical optimization procedure \eqref{eq:U-IDCOMain2} at the heart of U--SIDCO, that can get stuck in a bad local minimum. Figure \ref{fig:frames15} shows the coherences reached by sparse incoherent frames for a fixed $\ell_1$ regularization of $\lambda = 1.8$. Different $\lambda$s lead to frames with different coherence values and sparsity levels. The percentages indicate the overall sparsity levels. Observe that we do encounter a $10-20\%$ increase in coherence with the benefit of having approximately $50\%$ of the frame coefficients set to zero. Notice that the performance of the sparse complex frames is similar to that of the full real frames. In the case of sparse unital frames, we fix the zero entries a priori and optimize only over the rest of the entries. The zero entries are set randomly for each frame vector individually thus ensuring that they have the same sparsity level. We have chosen this approach for the sparse unital frame since adding the $\ell_1$ penalty to \eqref{eq:U-IDCOMain2} cannot produce any sparse solution (due to the conflict with the unit magnitude constraints).

The last test concerns nonnegative frames and the results are depicted in Figure \ref{fig:frames15}. In these cases the coherence levels are the highest, especially for the real frames. All these SIDCO derived methods run for $K = 2000$ iterations.

As expected, the harmonic and real frames provide the highest coherence values, since they have the fewest degrees of freedom. Interestingly, for high redundancy the performance gap between the two diminishes, and in some cases the harmonic frames achieve lower coherence than their real counterparts. Also, notice that for larger $m$, harmonic frames may approach the performance of general unital complex frames.

\begin{figure}[t]
\centering
\begin{subfigure}{0.42\textwidth}
\includegraphics[trim = 17 8 30 20, clip, width=\textwidth]{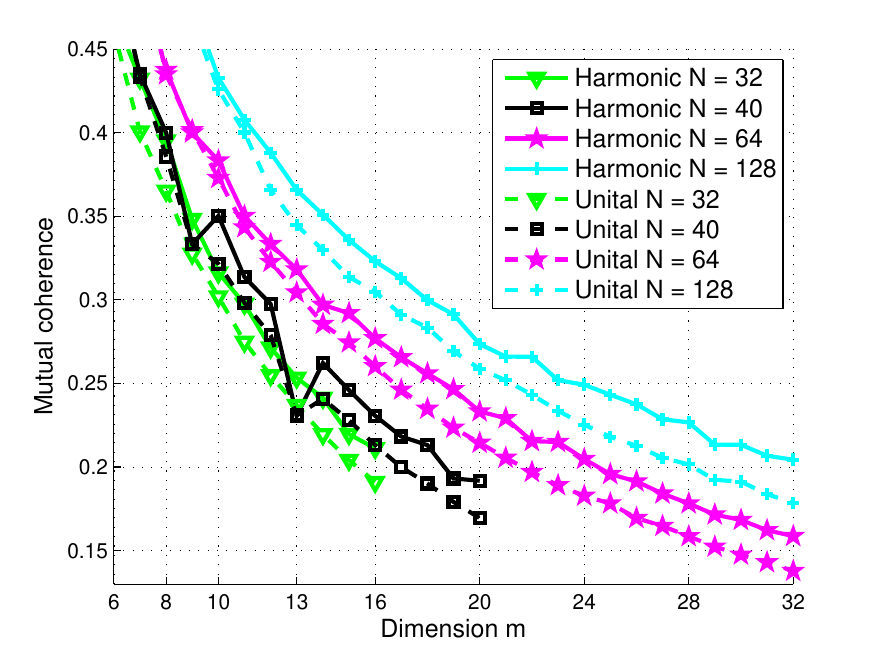}
\label{fig:wl1_fourier_design}
\end{subfigure}
\begin{subfigure}{0.42\textwidth}
\includegraphics[trim = 17 8 30 20, clip, width=\textwidth]{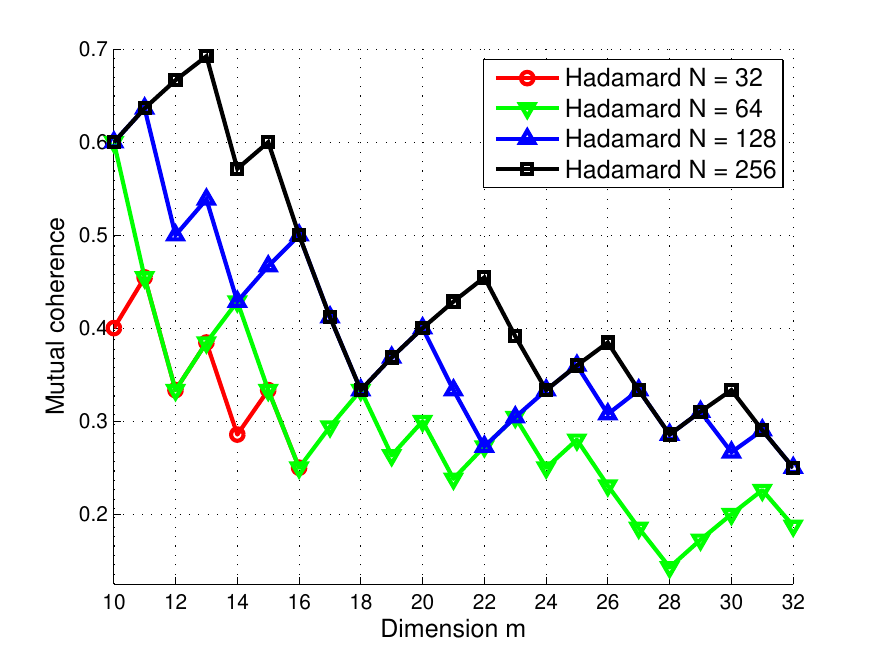}
\label{fig:wl1_hadamard_design}
\end{subfigure}
\caption{Fourier and Hadamard incoherent frames. Plot on the left: mutual coherence of harmonic frames $(m, N)$ designed via Algorithm 2. For $N = 32$ and all $m$ the lowest possible coherences are reached (this was checked by exhaustive search) and we also reach the known harmonic ETF $(13, 40)$. We show the frames up to $m = N/2$ and for $N \in \{ 64, 128 \}$ we show results up to $m=32$. We also show the results of U--SIDCO that performs better than the harmonic frames found; Plot on the right: mutual coherence of Hadamard frames $(m, N)$. For $N = 32$ the lowest possible coherences are reached (this was checked by exhaustive search). We show the frames up to $m = N/2$ and for $N \in \{ 128, 256 \}$ only up to $m=32$. The regularization parameter in \eqref{eq:ConvexHarmonicETF} is always $\lambda = m^{-1}$.}
\label{fig:wl1}
\end{figure}

Figure \ref{fig:wl1} shows the mutual coherence for several harmonic frames designed via Algorithm 2. For each frame $(m, N)$ we perform 500 runs. Also, we perform the same number of runs for the complementary frame $(N-m, N)$ and we choose the solution that provides the overall lowest mutual coherence. For $N = 32$ and all $m$, Algorithm 2 reaches the most incoherent harmonic frames (which was checked by exhaustive search). In the case $(13, 40)$ we do reach the known ETF. In all runs we set $K = 7, \lambda = m^{-1}$ and $\zeta = 0.1$. The local search parameter is $\ell = 4$ for $N \leq 40$, $\ell = 3$ for $N=64$ and finally $\ell = 2$ for $N = 128$. The proposed method seems to provide reasonable good solutions, especially given the computational complexity. Still, U--SIDCO performs better in most situations for $K = 300$ iterations (in a few it matches the incoherence of the harmonic frame of same dimension). We treat the harmonic frames separately in this test, because for a fixed choice of $(m,N)$ the computational complexity of exhaustively checking for the best (most incoherent) frames is impractical even on modern computing systems, and even for relatively small dimensions of $m$ and $N$. The gap to the best possible results offers a perspective on the performance of the proposed numerical solutions. The authors of \cite{CSWithFourier} design harmonic frames $(19, 381)$ with coherence $0.2820$ and $(29, 840)$, with coherence $0.1857$ using number theory tools. Algorithm 2 is not able to produce harmonic frames with lower coherences in these examples, while U--SIDCO is able to slightly improve the coherence to $0.2816$ for the frame $(19, 381)$.

The last simulations involve harmonic and real unital frames obtained from Hadamard matrices and the results are depicted in Figure \ref{fig:wl1}. The results are promising for the chosen dimensions, but exhaustive search is not practical for $N>32$. We do have performance references for dimensions where real unital ETFs are allowed and Hadamard matrices exist. We run the proposed algorithm for these dimensions and the results for $N \leq 1000$ are as follows: $(6, 16)$ the WB is reached, $(13, 40)$ we reach 0.5385 and the WB is 0.2308 which is attainable with a harmonic frame, $(28, 64)$ the WB is reached, $(20, 96)$ we reach 0.4 and the WB is 0.2, $(54, 160)$ we reach 0.1852 and the WB is 0.1111, $(120, 256)$ we reach 0.1 and the WB is 0.0667, $( 88, 320)$ we reach 0.1591 and the WB is 0.0909, $(72, 640)$ we reach 0.2222 and the WB is 0.1111 and finally for $(118, 768)$ we reach 0.1695 and the WB is 0.0847. These results show the effectiveness of the method. In the case $(28, 64)$ the ETF is reached by taking rows of the Hadamard matrix indexed in
$\mathcal{K} = \{4,     5,     6,    11,    13,    14,    16,    21,    23,    24,    25,    28,  32,    38,    39,  41,    42,   45,   48,   \dots \\    49,   50,    51,  53,    54,    55,    57,    61,    63 \}$.
Notice that both in the Fourier and Hadamard cases the mutual coherence is not monotonically decreasing with $m$. This is not surprising if we consider for example that when a difference set exists for $(m,N)$ then in general we do not also have a difference set for $(m+1, N)$ and therefore we will, most likely, have a temporary increase in coherence (see for example in Figure \ref{fig:wl1} the fall in coherence for the Harmonic ETF $(13, 40)$ and the Hadamard ETF $(28, 64)$). This phenomenon is exacerbated in the Hadamard case most probably due to the hard restriction on the entries of the Hadamard matrix. This is not the case for general (real or complex valued) frames where given a unit norm frame $\mathbf{F}_0 \in \mathbb{C}^{m \times N}$ we can always construct a new unit norm frame $\mathbf{F} \in \mathbb{C}^{(m+1) \times N}$ such that we at least have $\mu(\mathbf{F}) = \mu(\mathbf{F}_0)$ just by adding a zero row to $\mathbf{F}_0$.

\begin{table}[t]
\begin{center}
\caption{Coherence comparison of C--SIDCO against previously known methods following \cite{DifferenceSets}. Best results are underlined.}\label{tb:Comp1}
\begin{tabular}{|c|c|c|c|c|c|c|}
\hline
$(m,N)$ & C--SIDCO & \cite{BCASC} & \cite{Medra} & \cite{Love} & \cite{DifferenceSets} & CB \\ \hline
$(2,8)$ & \underline{0.7941} & 0.7950 & 0.7997 & 0.8415 & 0.8216 & 0.7500 \\ \hline
$(3, 16)$ & \underline{0.6486} & 0.6491 & 0.6590 & 0.8079 & 0.6766 & 0.6202 \\ \hline
$(4, 16)$ & \underline{0.4472} & \underline{0.4472} & 0.4473 & 0.7525 & 0.4514 & 0.4472 \\ \hline
$(4, 64)$ & 0.6906 & \underline{0.6869} & 0.7151 & 0.7973 & 0.7447 & 0.6000 \\ \hline
\end{tabular}
\end{center}
\end{table}
\begin{table}[t]
\begin{center}
\caption{Coherence comparison of C--SIDCO against previously known methods following \cite{Medra}. Best results are underlined.}\label{tb:Comp2}
\begin{tabular}{|c|c|c|c|c|c|}
\hline
$(m,N)$ & C--SIDCO & \cite{BCASC} & \cite{Medra} & \cite{Dhilon} & CB \\ \hline
$(4,6)$ & \underline{0.3273} & 0.3277 & 0.3274 & 0.3275 & 0.3162 \\ \hline
$(4,7)$ & \underline{0.3536} & \underline{0.3536} & 0.3540 & \underline{0.3536} & 0.3536 \\ \hline
$(4,8)$ & \underline{0.3780} & \underline{0.3780} & 0.3787 & 0.3782 & 0.3780 \\ \hline
$(4,9)$ & \underline{0.4021} & 0.4022 & \underline{0.4021} & 0.4034 & 0.3953 \\ \hline
$(4,10)$ & \underline{0.4113} & 0.4118 & \underline{0.4113} & 0.4114 & 0.4082 \\ \hline
$(4,20)$ & \underline{0.5000} & \underline{0.5000} & 0.5001 & 0.5335 & 0.5000 \\ \hline \hline
$(5,7)$ & \underline{0.2664} & 0.2670 & 0.2665 & 0.2669 & 0.2582 \\ \hline
$(5,8)$ & \underline{0.2952} & 0.2955 & 0.2954 & 0.2955 & 0.2928 \\ \hline
$(5,9)$ & \underline{0.3201} & 0.3207 & 0.3203 & 0.3216 & 0.3162 \\ \hline
$(5,10)$ & \underline{0.3333} & \underline{0.3333} & 0.3341 & 0.3336 & 0.3333 \\ \hline
$(5,16)$ & \underline{0.3889} & \underline{0.3889} & 0.3932 & 0.3959 & 0.3830 \\ \hline
\end{tabular}
\end{center}
\end{table}
A recent result has shown that, in compressed sensing applications, tight frames minimize the expected mean squared error \cite{WeiUNTF2012}, outperforming frames designed only with the incoherence target. Obviously, harmonic and Hadamard frames (constructed by Algorithm 2) are tight, so we are interested to check the frames constructed by Algorithm 1. General real and complex frames designed with the proposed algorithms approach the frame potential \cite{FP}, i.e., $\text{FP}(\mathbf{H}) = \| \mathbf{H}^H \mathbf{H} \|_F^2$, minimum value of $N^2/m$ that is known to be reached by tight frames (they are on average within $1\%$ of this bound). Unfortunately, in all other cases the frames, although highly incoherent, are no longer tight. In these cases we observe that $\mathbf{HH}^H$ is no longer exactly $N/m\mathbf{I}$ -- as mentioned, Grassmannian frames may not be tight frames in general. The unit polar decomposition can be applied on the final frames in the real, complex and unital cases (followed by an appropriate normalization) to lead to a tighter frame, with the cost of increasing the coherence. In the sparse and nonnegative cases, the polar decomposition may destroy the entire frame structure and thus must be avoided. If we have a redundant frame $\mathbf{H}$ with coherence $\gamma \mu$, where $\mu$ is the WB, and $\gamma \geq 1$ we have that $\text{FP}(\mathbf{H}) \leq \frac{N^2}{m} \left( \gamma^2 -  \frac{\gamma^2-1}{\rho} \right)$,
which shows that frames with coherence approaching the WB, i.e., $\gamma \approx 1$, have a bounded frame potential, close to the minimal bound of tight frames $N^2/m$.

\begin{table}[t]
\begin{center}
\caption{Coherence comparison of C--SIDCO and Algorithm 2 against the group theoretic and random constructions \cite{NewGroup}, respectively, when this approach does not reach ETFs. Best results are underlined.}\label{tb:Comp3}
\begin{tabular}{|c|c|c||c|c|c|} \hline
$(m,N)$ & \pbox{20cm}{\ \ Random \\ Fourier \cite{NewGroup}} & Algorithm 2 & \pbox{20cm}{\ \ \ Group \\ matrix \cite{NewGroup}} & C--SIDCO & WB \\ \hline
$(166, 499)$ & .1786 & \underline{.0949} & .0888 & \underline{.0649} & .0635 \\ \hline
$(260, 521)$ & .1504 & \underline{.0658} & .0458 & \underline{.0447} & .0439 \\ \hline
$(130, 521)$ & .2376 & \underline{.1190} & .1175 & \underline{.0796} & .0761 \\ \hline
$(214, 643)$ & .1978 & \underline{.0865} & .0755 & \underline{.0582} & .0559 \\ \hline
$(175, 701)$ & .2316 & \underline{.1023} & \underline{.0687} & .0700 & .0655 \\ \hline
$(350, 701)$ & .1326 & \underline{.0582} & .0393 & \underline{.0388} & .0379 \\ \hline
$(504, 1009)$ & .1147 & \underline{.0490} & \underline{.0325} & \underline{.0325} & .0315 \\ \hline
$(336, 1009)$ & .1384 & \underline{.0691} & .0597 & \underline{.0476} & .0446 \\ \hline
$(252, 1009)$ & .1631 & \underline{.0872} & .0846 & \underline{.0599} & .0546 \\ \hline
\end{tabular}
\end{center}
\end{table}
\subsection{Comparisons of C--SIDCO against previous methods}

We compare the proposed method C--SIDCO with previously, well-known, methods from the literature. Comparisons with the other SIDCO types of algorithms are not possible since previous methods are not able to accommodate additional constraints like unit magnitude entries or sparsity.

The results are presented in Tables \ref{tb:Comp1}, \ref{tb:Comp2} and \ref{tb:Comp3}, where we replicate and compare against the best previously known results. C--SIDCO runs for $K = 2000$ iterations and we show the best results out of 10 runs that are made with random initial frames. As can be seen from the tables, C--SIDCO provides the best results in all situations except one where it provides the second best result. We show the composite bound (CB) for coherence as defined in \cite{BCASC}. 
There are no large performance gaps since all methods perform quite well for the relative small dimensions chosen $(m, N)$. A recent result \cite{All3} has shown that the $(3,8)$ ETF, with WB $0.488$, cannot be constructed -- here C--SIDCO is able to achieve coherence 0.5, lower than the 0.6407 from \cite{Love}.

\subsection{Application of incoherent frames for sparse recovery}

\begin{figure}[t]
	\centering
	\begin{subfigure}{0.42\textwidth}
		\includegraphics[trim = 5 0 30 20, clip, width=\textwidth]{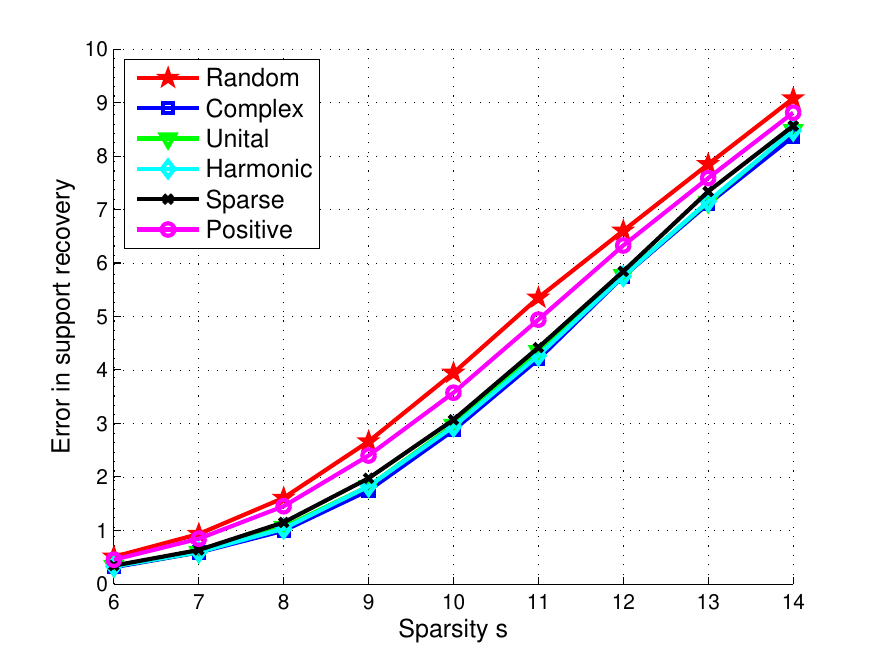}
	\end{subfigure}
	\centering
	\begin{subfigure}{0.42\textwidth}
		\includegraphics[trim = 5 0 30 20, clip, width=\textwidth]{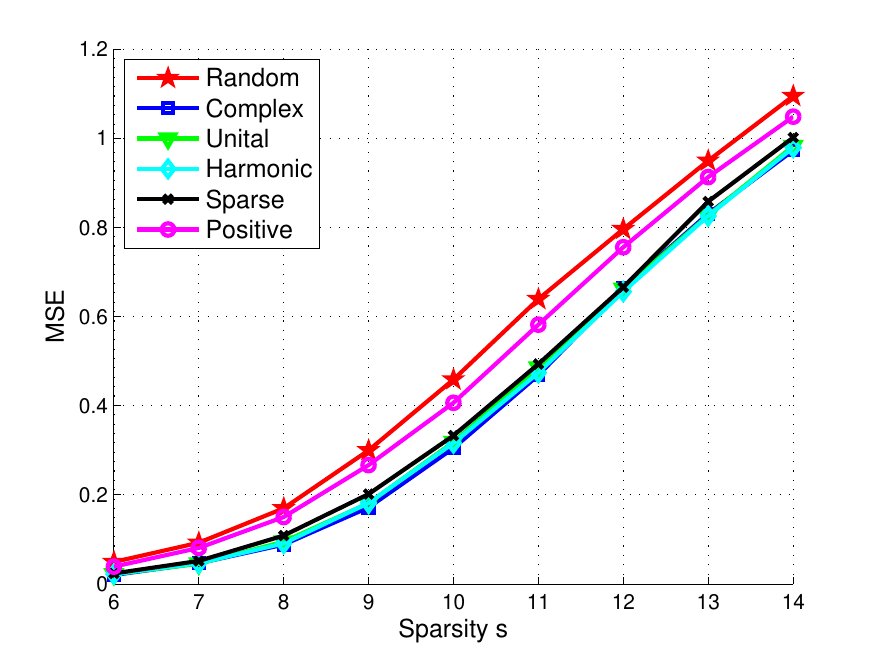}
	\end{subfigure}
	\caption{Sparse recovery performance, error in support accuracy (left) and mean-squared error (right), for incoherent frames of size $(25, 150)$ designed by the methods proposed in this paper as compared to a random frame. Recovery is done via OMP and the SNR level is $15$dB.}
	\label{fig:sparserecovery}
\end{figure}
In this section, we show the sparse recovery performance of the incoherent frames designed with the methods presented in this paper and we compare against random frames. We recover $s$-sparse vectors $\mathbf{x} \in \mathbb{C}^N$ from $m$ linear measurements $\mathbf{y} = \mathbf{Ax} + \mathbf{n}$ where $\mathbf{n} \in \mathbb{C}^m$ is i.i.d. white Gaussian noise. We fix $m = 25$ and $N = 150$ while the sparse recovery step is performed using the orthogonal matching pursuit (OMP) algorithm \cite{GreedIsGood}. We call the recovered $s$-sparse solution $\mathbf{\tilde{x}}$. The sparse vectors $\mathbf{x}$ are chosen randomly (the support is selected uniformly at random and the nonzero entries are drawn from the standard Gaussian distribution) and normalized such that $\| \mathbf{x} \|_2 = 1$.

Figure \ref{fig:sparserecovery} shows the average recovery results over $10^5$ realization of $\mathbf{x}$ with the incoherent measurement frames previously designed fixed. We show the average error in the support of the recovery $1/2(|\text{supp}(\mathbf{x}) \setminus \text{supp}(\mathbf{\tilde{x}})| + |\text{supp}(\mathbf{\tilde{x}}) \setminus \text{supp}(\mathbf{x})|) $, where $\text{supp}(\mathbf{z})$ returns the index set of the nonzero entries of $\mathbf{z}$, and the squared error $\|\mathbf{x} - \mathbf{\tilde{x}} \|_2^2$. When the sparsity $s$ is low all frames perform similarly while with higher $s$ the frames that do not have low coherence perform worse (the random frame performs worst followed by the positive frame designed via CP-SIDCO which has coherence $0.3233$). The coherence values are $0.1993$ for the general complex, $0.2268$ for the unital (with $\gamma = 0.01$ in \eqref{eq:U-IDCOMain2}), $0.2536$ for the harmonic (with $\lambda = 0.04$ in \eqref{eq:ConvexHarmonicETF}) and $0.2437$ for the sparse (with $\lambda = 1.8$ in \eqref{eq:SC-IDCOMain2} the frame has 54.52\% zero entries). These frames perform similarly across the sparsity $s$. The random frame is normalized such that its squared Frobenius norm is $N$, the same as the proposed incoherent frames.

\section{Conclusions}\label{sec:Conclusions}

In this manuscript, we introduce algorithms based on convex optimization for the design of highly incoherent real and complex frames under several constraints: nonnegativity, sparsity and unit magnitude. We design highly incoherent frames for every dimension and under the constraints previously enumerated. We deal with two cases: designing general frames and frames from rows of fixed known unital matrices (like Fourier and Hadamard). We show the results of the proposed methods relative to the performance limit of the Welch bound and that in the general complex case the proposed methods match or outperform previously proposed methods.

\section*{References}
\bibliographystyle{IEEEtran}
\bibliography{refs}

\end{document}